\documentclass[fleqn,usenatbib]{mnras}

\usepackage{newtxtext,newtxmath}
\usepackage[T1]{fontenc}

\DeclareRobustCommand{\VAN}[3]{#2}
\let\VANthebibliography\thebibliography
\def\thebibliography{\DeclareRobustCommand{\VAN}[3]{##3}\VANthebibliography}

\usepackage{graphicx}	% Including figure files
\usepackage{amsmath}	% Advanced maths commands
\usepackage{float}
\usepackage{multirow}
\usepackage{ulem}
\usepackage{listings}
\usepackage{xcolor}
\usepackage{hyperref}
\usepackage[symbol]{footmisc}
\definecolor{lightgray}{gray}{0.9}

\newcommand{\packname}{{\tt PromoPlot}}

\graphicspath{{./}{figures/}}
\title[Promo Plots]{\fontsize{18}{22}\selectfont PromoPlot: Covering open-access fees by filling wasted space in corner plots}

\author[Rowan and Roberts]{%
Dominick M. Rowan$^{1,2}$ and John D. Roberts$^{1,2}$
 \\
$^{1}$Department of Astronomy, The Ohio State University, 140 W. 18th Ave., Columbus, OH 43210, USA\\
$^{2}$Center for Cosmology and Astroparticle Physics, The Ohio State University, Street Address, Columbus, OH 43210, USA\\
}

\date{Accepted XXX. Received 2025-04-01; in original form 2025-04-01}

\pubyear{2025}

\begin{document}
\label{firstpage}
\pagerange{\pageref{firstpage}--\pageref{lastpage}}
\maketitle

\begin{abstract}

In an effort to reduce drain on grant funds and decrease unused space in publications, we have developed a Python package for inserting advertisements into the space left empty by corner plots. This novel technique can allow authors to reduce or eliminate publication charges for journals such as MNRAS and ApJ. In order to offset publication costs entirely, we recommend that authors include anywhere from 200-700 corner plots per paper, with the exact number depending on the journal used. Finally, we discuss other opportunities to generate ad revenue through publications.

\end{abstract}

\begin{keywords}

Publications (Several) --- Standards (High) --- Economics (1) 
\end{keywords}

%%%%%%%%%%%%%%%%%%%%%%%%%%%%%%%%%%%%%%%%%%%%%%%%%%

%%%%%%%%%%%%%%%%% BODY OF PAPER %%%%%%%%%%%%%%%%%%

\section{Introduction} \label{sec:intro}

The ascension of the open-access journal has been accompanied by increasing dependence on author publication fees. While this transition has strong merits in ensuring a wide dissemination of research, without subscription fees, publishers now rely on authors to cover the journal operating costs. As of last year, nearly 50\% of all papers are published in open-access or hybrid open-access journals where the author fronts the majority of the bill \citep{Brainard24}. In 2022, all journals published through the American Astronomical Society (AAS) switched to an open-access model. In 2024, the Monthly Notices of the Royal Astronomy Society (MNRAS) followed suit. While there are some journals, like the Open Journal of Astronomy, that are peer-reviewed, open-access, and do not have author publication costs, such journals are the clear exception. 

Open-access journals promote broader access to research and can increase visibility and impact, both laudable outcomes. Purely coincidentally, they have also served as effective sources of revenue for publishers. Between 2019 and 2023, \citet{Haustein24} found that the spending by authors on open-access fees nearly tripled, generating \$8.97 billion over the five-year period for a selection of six publishers. 

%In addition, by passing the costs to the authors, the ability to publish in high-impact journals becomes linked to the affluence of the author's country or institution. While some journals, including the AAS Journals \footnote{\url{https://journals.aas.org/support}} and MNRAS\footnote{\footnotesize\url{https://academic.oup.com/pages/purchasing/developing-countries-initiative/}}, offer waivers or discounts to scientists in low-income countries, it has been shown that these can still not be enough to incentive scientists to publish in open-access journals \citep{Momeni23}.

Publication costs can range from \$1500-\$5000 per article, detracting from funds that can be used for conducting research, hiring students, and participating in conferences. It is unlikely that the open-access policies or author payment costs will change in the near future\footnote{We would consider asking them to lower such fees, but that would require making a phone call, which is beyond the scope of this paper.}. We therefore are incentivized to identify opportunities to offset these costs. We can start by looking at sources of revenue generation used by other online media platforms that can be applied to scientific publications. The most common business model used online is advertisements. Entire services and website development packages have been built to seamlessly integrate advertisements on websites. In online media, ads are commonplace, appearing as banners or sidebars, pop-ups, or at bookends of online videos. 

The most successful advertisements are carefully placed within the content so that they do not detract from the website or media. For example, reading a paper that used in-line advertisements which push the text around would be as unpleasant as trying to find the recipe on a cooking blog. Such situations must be avoided at all costs. Instead, our goal is to identify and fill space that is currently wasted in most manuscripts and utilize that space to generate ad revenue. 

The availability of open-source sampling codes, such as {\tt emcee} \citep{ForemanMackey13} and {\tt DYNESTY} \citep{Speagle20}, has made parameter estimation for complex, multimodal distributions commonplace in the literature. As a result, studies using these methods often present posteriors in the form of a ``triangle'' or ``corner'' plot. These are typically structured with histograms on the diagonal and the covariances for each parameter shown above or below the diagonal. These figures provide a straightforward way to view results of sampling methods, but they also render a substantial part of the page unusable, wasting valuable page costs. 

\begin{figure*}
    \centering
    \includegraphics[width=\linewidth]{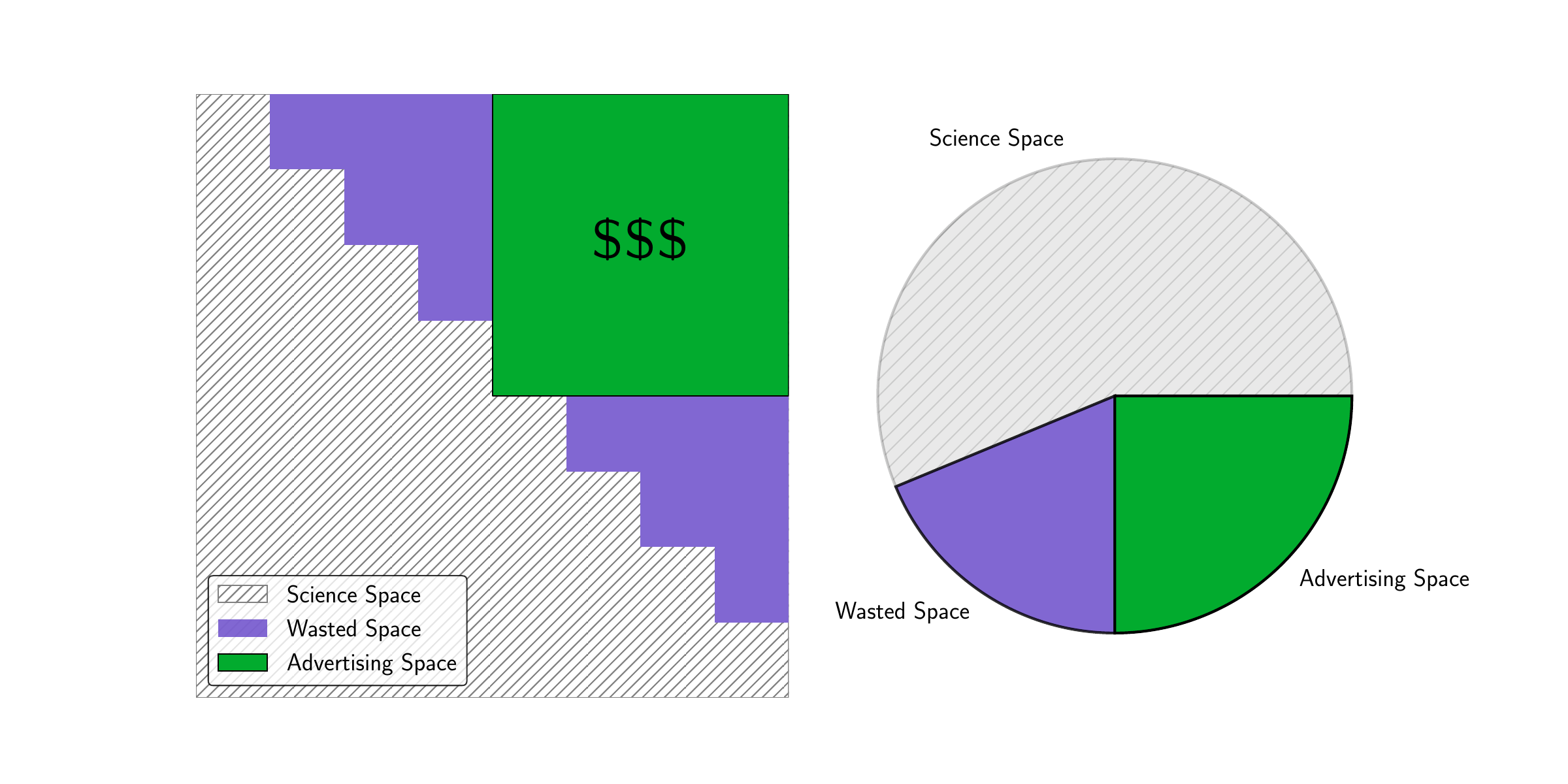}
    \caption{With \packname{}, the amount of space wasted with corner plots can be reduced substantially. In this example, $\sim 25\%$ of the figure is now dedicated to advertising. This results a $\sim 57\%$ reduction in wasted space for this figure as well.}
    \label{fig:saving_space}
\end{figure*}

The prevalence of such figures is only increasing. As of April 1st, 2025, more than 1,700 papers cite the {\tt corner} package \citet{ForemanMackey16}, with more than 350 of these citations coming from 2024 alone. There have been approximately 400 papers citing {\tt corner} in both ApJ and MNRAS spanning a range of fields including exoplanets \citep{Jones24}, cosmology \citep{Turner24}, stars \citep{Rowan25}, and the interstellar medium \citep{Sun18}. In some cases, $\gtrsim 10$ corner plots are used in a single publication \citep[e.g.,][]{Calabrese2025}. 

We therefore propose to solve two problems -- page charges and wasted space in corner plots -- with a single solution by selling ad space in the upper half of corner plots. In Section \S\ref{sec:Implementation}, we present our Python package, \packname{}, and demonstrate how users can automatically include advertisements in their corner plots. In Section \S\ref{sec:profits}, we determine how much the use of ads in corner plots can reduce page charges. Finally, we describe the prospects for wide adoption of this practice in the field, and the possibilities for selling targeted ads.

\section{Implementation of Ads in Plots} \label{sec:Implementation}

\begin{figure*}
    \centering
    \href{https://monday.com}{\includegraphics[width=0.8\linewidth]{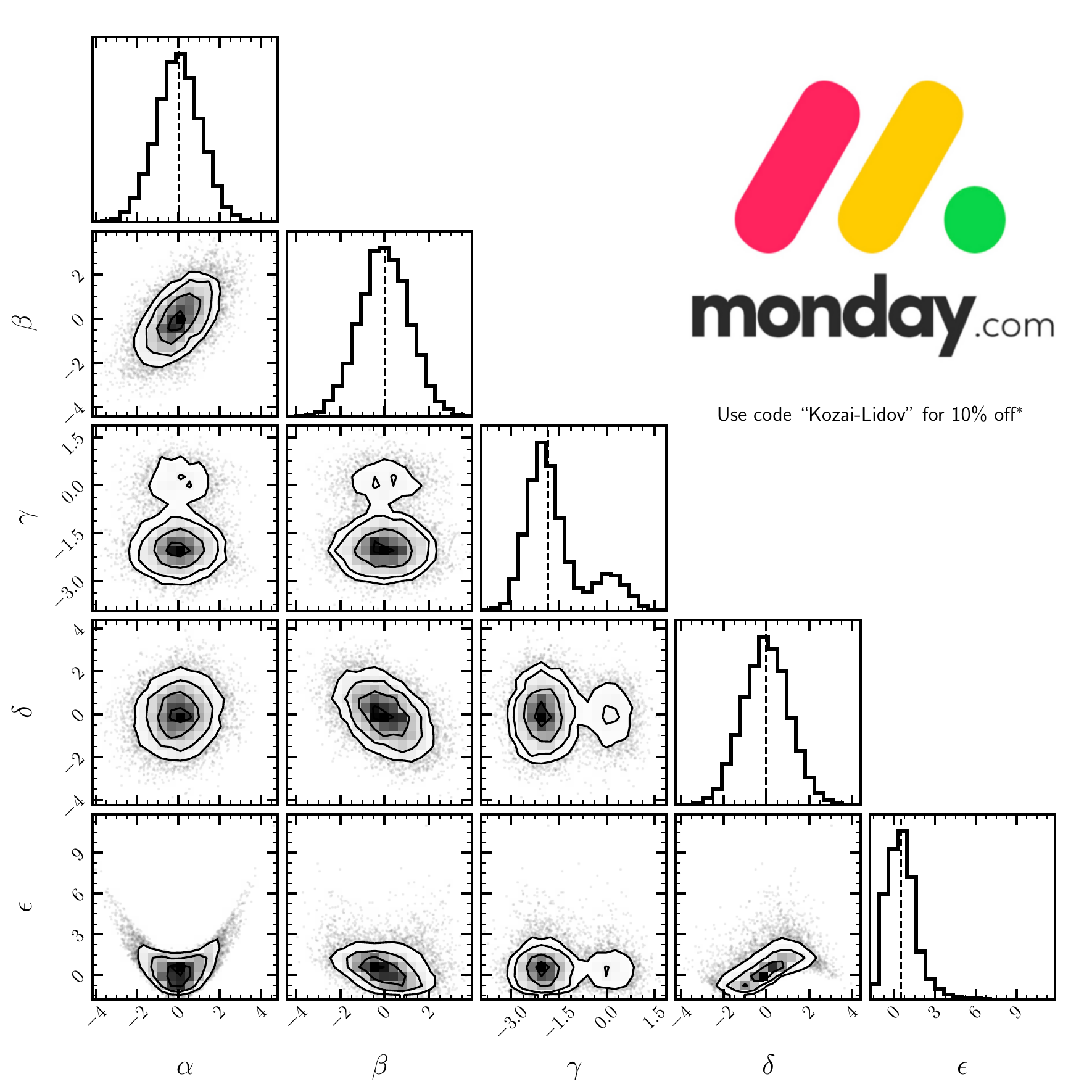}}
    \caption{Previously, the upper half of the plot would be totally wasted. With \packname{}, we can fill the space and generate revenue through advertisements.}
    \small\textsuperscript{*} Promotional codes are included for demonstration purposes and may not provide listed discount.
    \label{fig:example_corner_equal}
\end{figure*}

\begin{figure*}
    \centering
    \href{https://blueapron.com}{\includegraphics[width=0.8\linewidth]{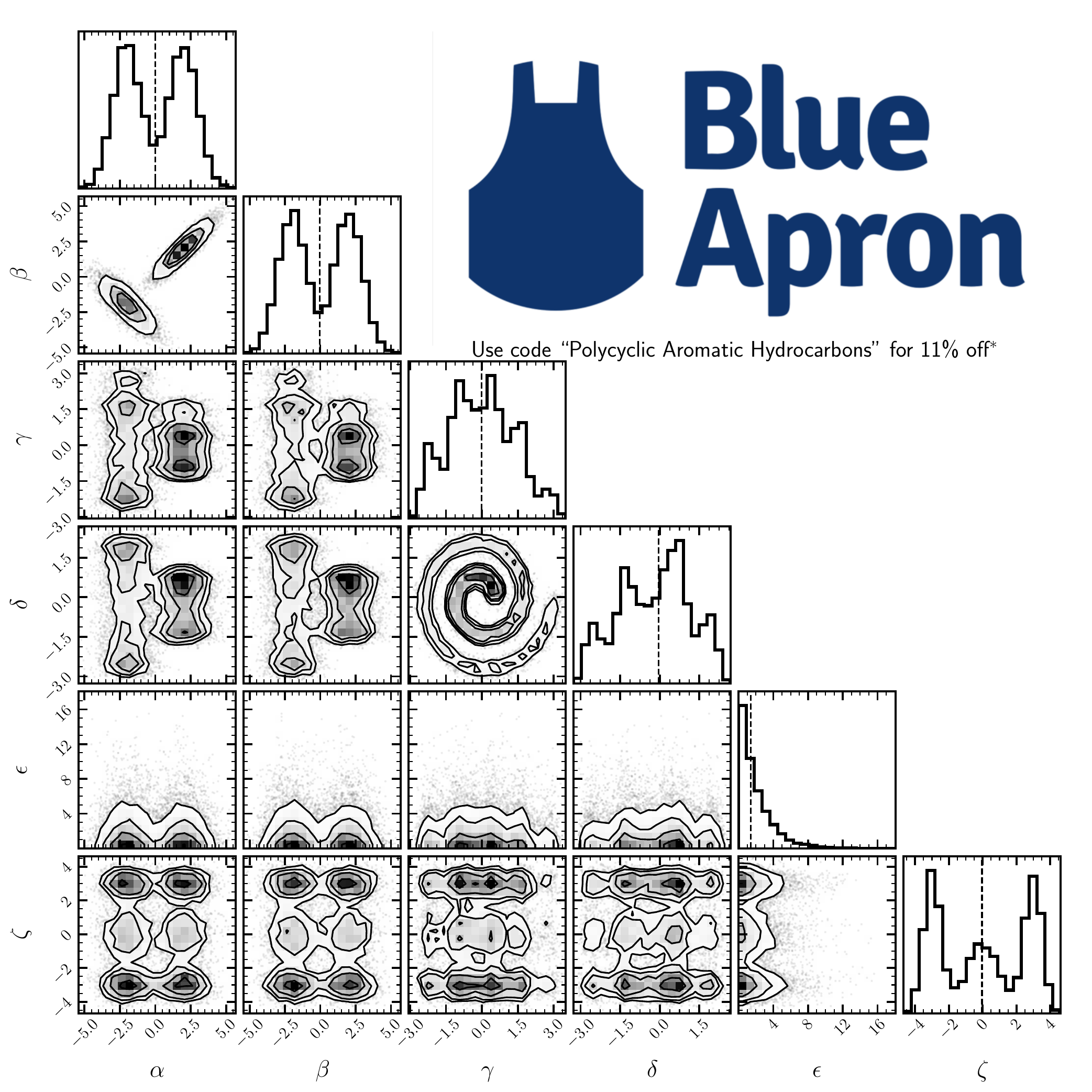}}
    \caption{A second example of advertisements with \packname{}.}
    \small\textsuperscript{*} Promotional codes are included for demonstration purposes and may not provide listed discount.
    \label{fig:example_corner_wide}
\end{figure*}

The \packname{} package is available as an open-source code at \url{https://github.com/dmrowan/promoplot}. The package is built on {\tt corner} \citep{ForemanMackey16}, and includes a number of advertising partners pre-installed. Updating your corner plots to include promotional material can be done by modifying a single line of code:

\begin{lstlisting}
from promoplot import PromoPlot

df = #your data here

kwargs = dict(
        hist_kwargs=dict(lw=3, color='black'),
        labels = df.columns,
        quantiles=[0.5],
        label_kwargs=dict(fontsize=22))

fig = PromoPlot(df, **kwargs)
\end{lstlisting}

It is also straightforward to include specific advertisers and promotional codes for special offers. Figure \ref{fig:saving_space} shows how \packname{} makes use of the available page real estate to maximize profits with a single, equal-aspect ratio advertisement. Figures \ref{fig:example_corner_equal} and \ref{fig:example_corner_wide} show two examples with mock data.

    %Coca-Cola
    %Pepsi
    %UberEats
    %Dunkin Donuts
    %NordVPN
    %Hello Fresh
    %Blue Apron
    %BetterHelp
    %Brilliant
    %Honey
    %Raid: Shadow Legends
    %SquareSpace (heh)
    %Draft Kings

\section{Profit Analysis}\label{sec:profits}

\begin{figure*}
    \centering
    \includegraphics[width=0.8\linewidth]{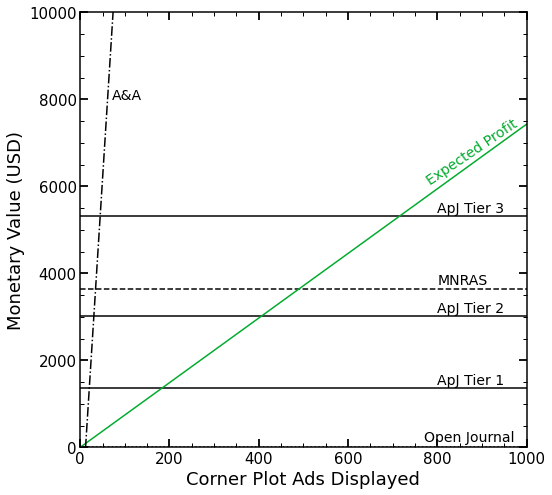}
    \caption{Comparison of expected profits to costs for different journals. As can be seen, the lack of page fees with most journals allows large numbers of corner plots to be added to recoup the entire cost. ApJ fees can be covered from between 183-716 corner plots. MNRAS can be covered by 490. Open Journal papers and A\&A papers less than 12 pages are free, which represent actual profit made for any corner plot. However, fees per page outpace earnings rapidly in A\&A publications.}
    \label{fig:profit}
\end{figure*}

Our negotiations with advertisers are still pending, so we do not have concrete numbers for the revenue generated per advertisement. Instead, we compare with the amount of revenue generated for the New York Times (NYT) in terms of ad space. To place an advertisement in the NYT that takes up one third of a page, advertisers would need to pay \$25,460. The value of this adspace comes from the large viewership of the NYT (estimated at 10 million) and the large amount of page real estate. We will use this as the basis of our estimates.

Estimated revenue will depend on the readership of the journal and the amount of space used for the ad. For this paper, we will assume a publication to the Astrophysical Journal (ApJ) and a full page figure corner plot with 7 plots. We assume ApJ has a readership of order 10,000 readers, and that every astronomer reads every paper. Additionally, the ad is smaller on the page, taking up about 9.73\% of the page for our example figure. With these assumptions, scaling down the costs appropriately gives an estimate of \$7.43 in revenue per corner plot. 

Figure \ref{fig:profit} shows comparisons of the expected profit and publishing fees for a variety of journals. As an example, a tier 2 paper published to ApJ (typical length paper), the page fees would amount of \$3011, which amounts to including 406 corner plot ads. While normally including 406 additional figures would bump into a higher bracket for paper fees, it is possible to include all the plots in an online \textit{figure set} to only minimally impact the paper size. We have not heard back from our advertisers about the relative ad prices for online figure sets, so we will assume they are the same. However, as can be seen, for most journals, this can be overcome by simply including more corner plots. The sole exception being A\&A, which does have a fee per page.

Based on these results \textbf{we recommend all future publications submitted to ApJ use at least 200 corner plots per paper. We also recommend all MNRAS publications include at least 490 corner plots.}

\section{Looking Forward}\label{sec:conclusion}

In this paper, we have presented a new tool to offset increasing publication costs with \packname{}. By taking advantage of currently unused space in manuscripts, we aim to reduce financial strain in the scientific process. \packname{} comes preinstalled with a number of advertisers, and we demonstrate how this tool can be seamlessly integrated into your Python workflow.\footnote{If you are still using {\tt supermongo}, you are on your own.}

\begin{figure*}
    \centering
    \href{https://duolingo.com}{\includegraphics[width=0.9\linewidth]{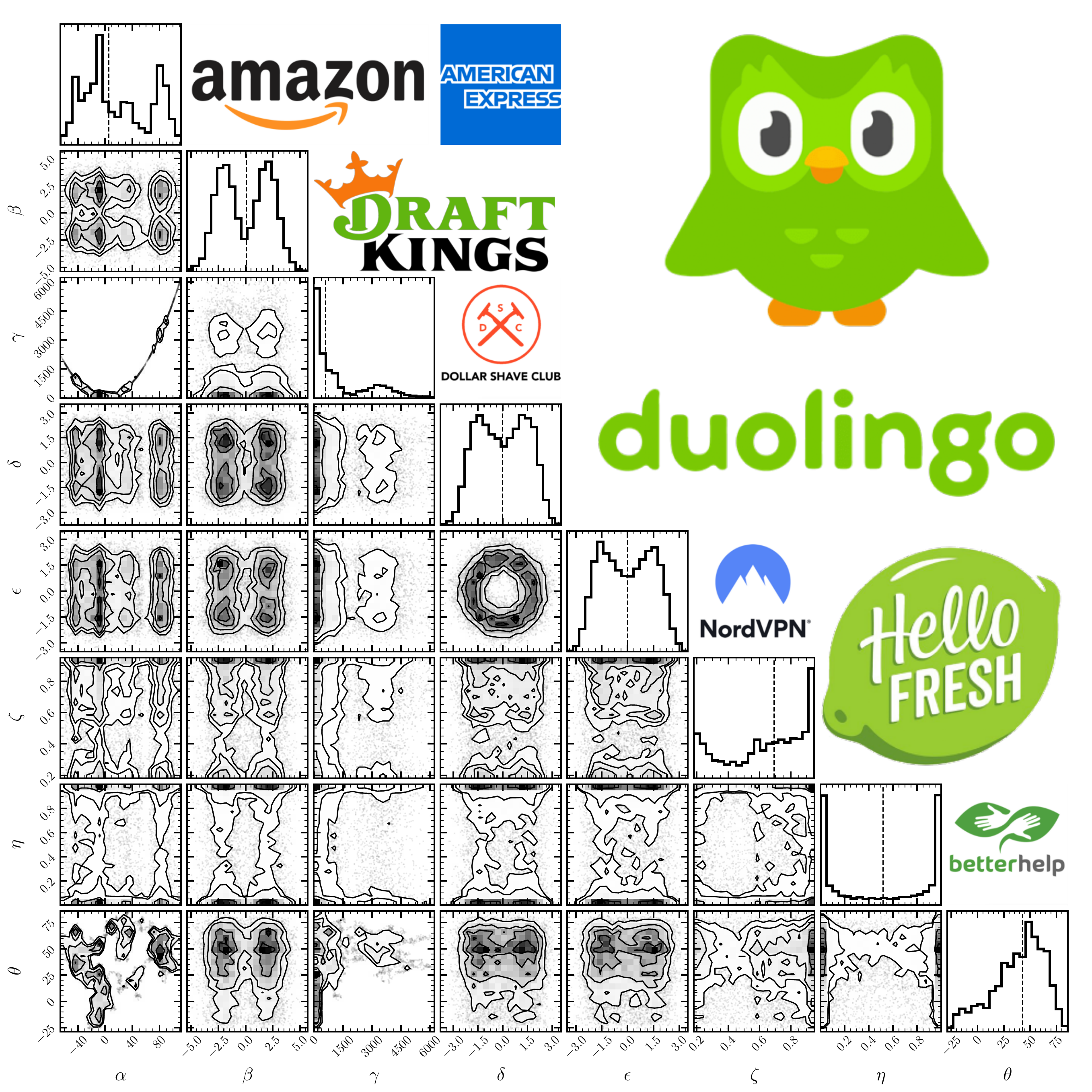}}
    \caption{Example of how multiple ads can be used to fill the entire space left empty.}
    \label{fig:multiple_ads}
\end{figure*}

The current version of \packname{} (v. 1.2.0) only includes one advertisement per figure. Given that the majority of advertisements are in 1:1 or 2:1 aspect ratio, there is still unused space available adjacent to the primary advertisement (Figure \ref{fig:saving_space}). We have preliminary progress towards fixing this issue as well. Figure \ref{fig:multiple_ads} shows an example of a development feature, which will be available in a future major release.

We realize that using \packname{} alone to offset publication costs entirely would require authors to include more corner plots than before.\footnote{Some of our colleagues have even suggested too many. We have been unable to verify this statement.} However, there are some other promising directions for expanding the profitability of papers. For instance, by partnering directly with advertisers, astronomers can provide discount codes in their \packname{} (e.g., Figure \ref{fig:example_corner_equal}) and receive kickbacks when they are used by their readers. We recommend that such partnerships be developed at the institutional or collaboration level to reduce burden on early-career scientists. 

%We have demonstrated that using \packname{} alone is unlikely to offset publication costs entirely for the majority of authors. There are some other promising directions for expanding the profitability of papers. For instance, by partnering directly with advertisers, astronomers can provide discount codes in their \packname{} (e.g., Figure \ref{fig:example_corner_equal}) and receive kickbacks when they are used by their readers. We recommend that such partnerships be developed at the institutional or collaboration level to reduce burden on early-career scientists. 

While general advertisements for widely-used products and services, like those presented here, are capable at targeting a broad audience, there are also exciting opportunities to insert dynamically generated ads targeted to the interests of individual readers. Targeted ads have been shown to increase revenue by a factor of 2.7 per ad \citep{BealesEisenach2014}, and increase the chance that the person who views that ad goes on to purchase the product or service. This practice would likely require adoption by journals and pre-print servers in order to embed advertisements when the article is accessed and downloaded. We will be reaching out to each journal soon to discuss possible methods of rolling out such features. 

\section*{Data Availability}

The package used to generate these plots is available at: \url{https://github.com/dmrowan/promoplot/}



\bibliographystyle{mnras}
\bibliography{biblio}

\begin{thebibliography}{}
\makeatletter
\relax
\def\mn@urlcharsother{\let\do\@makeother \do\$\do\&\do\#\do\^\do\_\do\%\do\~}
\def\mn@doi{\begingroup\mn@urlcharsother \@ifnextchar [ {\mn@doi@} {\mn@doi@[]}}
\def\mn@doi@[#1]#2{\def\@tempa{#1}\ifx\@tempa\@empty \href {http://dx.doi.org/#2} {doi:#2}\else \href {http://dx.doi.org/#2} {#1}\fi \endgroup}
\def\mn@eprint#1#2{\mn@eprint@#1:#2::\@nil}
\def\mn@eprint@arXiv#1{\href {http://arxiv.org/abs/#1} {{\tt arXiv:#1}}}
\def\mn@eprint@dblp#1{\href {http://dblp.uni-trier.de/rec/bibtex/#1.xml} {dblp:#1}}
\def\mn@eprint@#1:#2:#3:#4\@nil{\def\@tempa {#1}\def\@tempb {#2}\def\@tempc {#3}\ifx \@tempc \@empty \let \@tempc \@tempb \let \@tempb \@tempa \fi \ifx \@tempb \@empty \def\@tempb {arXiv}\fi \@ifundefined {mn@eprint@\@tempb}{\@tempb:\@tempc}{\expandafter \expandafter \csname mn@eprint@\@tempb\endcsname \expandafter{\@tempc}}}

\bibitem[\protect\citeauthoryear{Beales \& Eisenach}{Beales \& Eisenach}{2014}]{BealesEisenach2014}
Beales H.,  Eisenach J.~A.,  2014, \mn@doi [SSRN Electronic Journal] {10.2139/ssrn.2421405}

\bibitem[\protect\citeauthoryear{Brainard}{Brainard}{2024}]{Brainard24}
Brainard J.,  2024, \mn@doi [Science] {https://doi.org/10.1126/science.ads1013}, 385

\bibitem[\protect\citeauthoryear{{Calabrese} et~al.,}{{Calabrese} et~al.}{2025}]{Calabrese2025}
{Calabrese} E.,  et~al., 2025, \mn@doi [arXiv e-prints] {10.48550/arXiv.2503.14454}, \href {https://ui.adsabs.harvard.edu/abs/2025arXiv250314454C} {p. arXiv:2503.14454}

\bibitem[\protect\citeauthoryear{{Foreman-Mackey}}{{Foreman-Mackey}}{2016}]{ForemanMackey16}
{Foreman-Mackey} D.,  2016, \mn@doi [The Journal of Open Source Software] {10.21105/joss.00024}, \href {https://ui.adsabs.harvard.edu/abs/2016JOSS....1...24F} {1, 24}

\bibitem[\protect\citeauthoryear{{Foreman-Mackey}, {Hogg}, {Lang}  \& {Goodman}}{{Foreman-Mackey} et~al.}{2013}]{ForemanMackey13}
{Foreman-Mackey} D.,  {Hogg} D.~W.,  {Lang} D.,   {Goodman} J.,  2013, \mn@doi [\pasp] {10.1086/670067}, \href {https://ui.adsabs.harvard.edu/abs/2013PASP..125..306F} {125, 306}

\bibitem[\protect\citeauthoryear{{Haustein}, {Schares}, {Alperin}, {Hare}, {Butler}  \& {Sch{\"o}nfelder}}{{Haustein} et~al.}{2024}]{Haustein24}
{Haustein} S.,  {Schares} E.,  {Alperin} J.~P.,  {Hare} M.,  {Butler} L.-A.,   {Sch{\"o}nfelder} N.,  2024, \mn@doi [arXiv e-prints] {10.48550/arXiv.2407.16551}, \href {https://ui.adsabs.harvard.edu/abs/2024arXiv240716551H} {p. arXiv:2407.16551}

\bibitem[\protect\citeauthoryear{{Jones} et~al.,}{{Jones} et~al.}{2024}]{Jones24}
{Jones} S.~E.,  et~al., 2024, \mn@doi [\aj] {10.3847/1538-3881/ad55ea}, \href {https://ui.adsabs.harvard.edu/abs/2024AJ....168...93J} {168, 93}

\bibitem[\protect\citeauthoryear{{Rowan} et~al.,}{{Rowan} et~al.}{2025}]{Rowan25}
{Rowan} D.~M.,  et~al., 2025, \mn@doi [The Open Journal of Astrophysics] {10.33232/001c.129962}, \href {https://ui.adsabs.harvard.edu/abs/2025OJAp....8E..18R} {8, 18}

\bibitem[\protect\citeauthoryear{{Speagle}}{{Speagle}}{2020}]{Speagle20}
{Speagle} J.~S.,  2020, \mn@doi [\mnras] {10.1093/mnras/staa278}, \href {https://ui.adsabs.harvard.edu/abs/2020MNRAS.493.3132S} {493, 3132}

\bibitem[\protect\citeauthoryear{{Sun} et~al.,}{{Sun} et~al.}{2018}]{Sun18}
{Sun} J.,  et~al., 2018, \mn@doi [\apj] {10.3847/1538-4357/aac326}, \href {https://ui.adsabs.harvard.edu/abs/2018ApJ...860..172S} {860, 172}

\bibitem[\protect\citeauthoryear{{Turner} et~al.,}{{Turner} et~al.}{2024}]{Turner24}
{Turner} W.,  et~al., 2024, \mn@doi [\apj] {10.3847/1538-4357/ad8239}, \href {https://ui.adsabs.harvard.edu/abs/2024ApJ...976..143T} {976, 143}

\makeatother
\end{thebibliography}

%%%%%%%%%%%%%%%%% APPENDICES %%%%%%%%%%%%%%%%%%%%%

%\input{UnusedAppendix}

%%%%%%%%%%%%%%%%%%%%%%%%%%%%%%%%%%%%%%%%%%%%%%%%%%

% Don't change these lines
\bsp	% typesetting comment
\label{lastpage}
\end{document}